\documentclass{iau}

\usepackage{stmaryrd}
\usepackage{graphicx}
\usepackage{natbib}
\usepackage{enumitem}
\usepackage[breaklinks,colorlinks=true,citecolor=blue,linkcolor=blue,urlcolor=blue]{hyperref}

\newcommand{\apj}{\textit{ApJ}} 
\newcommand{\apjl}{\textit{ApJ}} 
\newcommand{\aap}{\textit{A\&A}} 
\newcommand{\mnras}{\textit{MNRAS}} 
\newcommand\pasj{\textit{PASJ}} 

\title[IAUS 345.~~Star formation in cloud cores] 
{Star formation in cloud cores -- simulations and observations of dense molecular cores and the formation of solar mass stars}

\author[Federrath]   
{C.~Federrath$^1$}

\affiliation{
$^1$Research School of Astronomy and Astrophysics, Australian National University, \\Canberra, ACT~2611, Australia \\ \vspace{0.1cm} email: {\tt christoph.federrath@anu.edu.au}
}

\pubyear{2019}
\volume{345}  
\jname{Origins: from the Protosun to the First Steps of Life}
\editors{Bruce G.~Elmegreen, L.~Viktor T\'oth, Manuel G\"udel, eds.}

\begin{document}

\maketitle

\begin{abstract}
Star formation is inefficient. Recent advances in numerical simulations and theoretical models of molecular clouds show that the combined effects of interstellar turbulence, magnetic fields and stellar feedback can explain the low efficiency of star formation. The star formation rate is highly sensitive to the driving mode of the turbulence. Solenoidal driving may be more important in the Central Molecular Zone, compared to more compressive driving agents in spiral-am clouds. Both theoretical and observational efforts are underway to determine the dominant driving mode of turbulence in different Galactic environments. New observations with ALMA, combined with other instruments such as CARMA, JCMT and the SMA begin to reveal the magnetic field structure of dense cores and protostellar disks, showing highly complex field geometries with ordered and turbulent field components. Such complex magnetic fields can give rise to a range of stellar masses and jet/outflow efficiencies in dense cores and protostellar accretion disks.
\end{abstract}

\firstsection

\section{The big picture}

\begin{figure*}
\centerline{\includegraphics[width=0.7\linewidth]{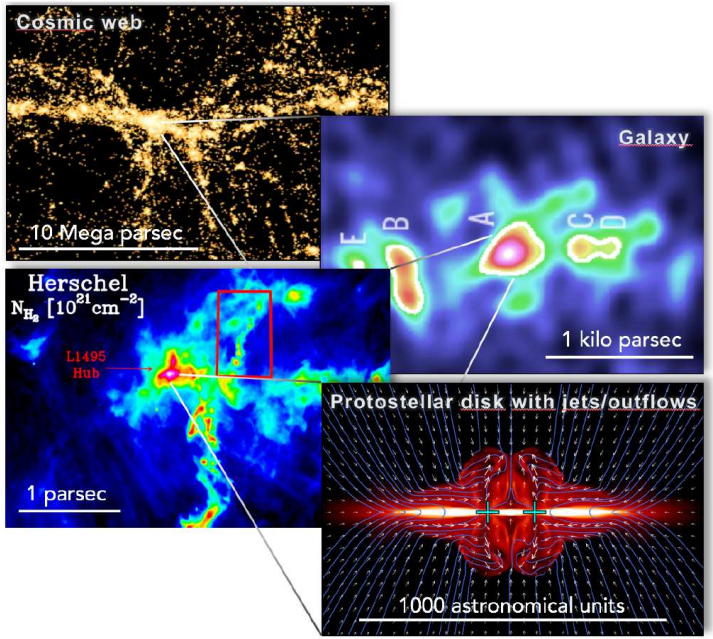}}
\caption{Schematic zoom-in on cosmic star formation. The spatial range covered by star formation is more than a billion: from the cosmic web (size scales of order 10 Mega parsec) to galaxies, interstellar clouds, and finally to the protostellar disks (scales of order 1000 astronomical units) that spawn new stars and planets. Observing, modeling, and understanding the complex interplay of the physical and chemical processes across this huge range of scales from the cosmic web down to stars and planets is one of the biggest challenges in astrophysics. Images adopted from \citet{TaylorKobayashi2015}, \citet{ShardaEtAl2018}, \citet{ArzoumanianEtAl2018}, and \citet{KuruwitaFederrathIreland2017}.}
\label{fig:cosmo}
\end{figure*}

The formation of stars powers the evolution of galaxies and determines the initial conditions for planet formation, and thus, ultimately for life. Star formation happens in the dense cores of filamentary molecular clouds, but many scales and processes are involved, ranging from cosmological initial conditions giving rise to the first galaxies, over the dense clouds within, to finally the protostellar accretion disks where binaries and planets form (see Figure~\ref{fig:cosmo}). This is an extremely rich physics problem, involving gravity, turbulence, magnetic fields, and feedback via jets, outflows and radiation. We are far from having solved all these problems, but new observations help us map the structure of the turbulent gas and magnetic fields in the dense protostellar cores, and new computer models are reaching the level of detail that enables a quantification of the star formation rate and the initial mass function of stars.

\section{The inefficiency of star formation}

Star formation is inefficient. Molecular clouds only turn a few percent of their gas mass into stars per freefall time. If gravity were the only thing acting on the clouds, we would expect the star formation rate to be of the order of 100\% per freefall time (i.e., the entire cloud would be turned into stars in one freefall time). Thus, physical processes other than gravity must be limiting the collapse and reducing the star formation rate to only a few percent per freefall time.

\begin{figure*}
\centerline{\includegraphics[width=0.95\linewidth]{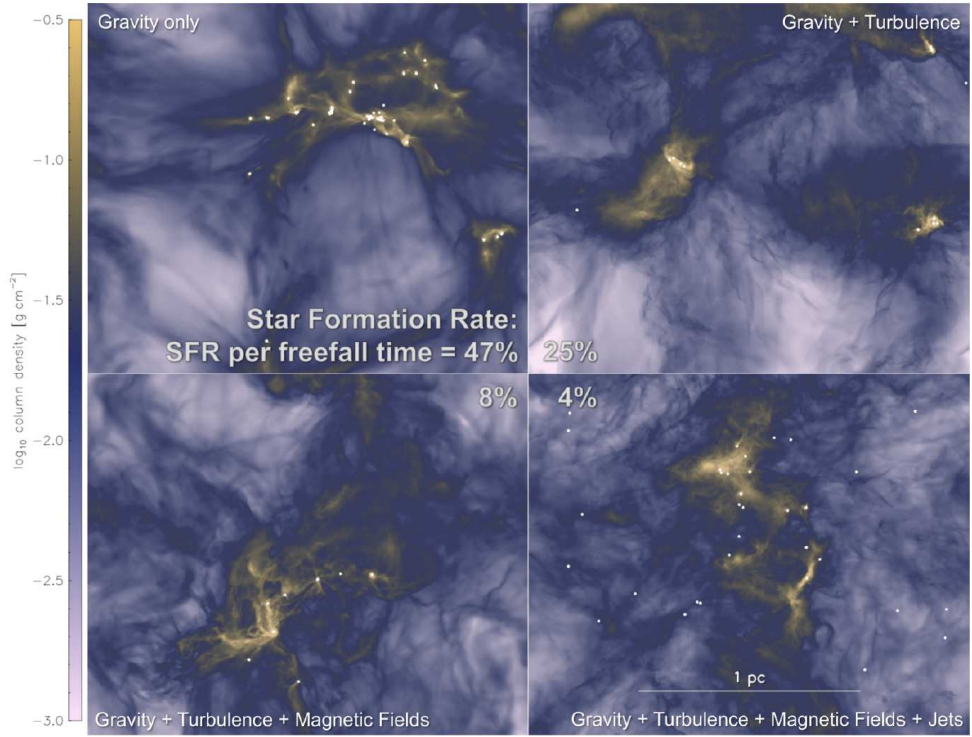}}
\caption{Four numerical simulations of star formation in the same clouds, but with systematically increasing physical complexity \citep{Federrath2015}: Gravity only (top left), Gravity vs.~Turbulence (top right), Gravity vs.~Turbulence + Magnetic Fields (bottom left), and Gravity vs.~Turbulence + Magnetic Fields + Jet/Outflow Feedback (bottom right). Stars are shown as white circles and the resulting star formation rates are indicated in each panel. Only the combination of gravity, turbulence, magnetic fields, and feedback yields realistic star formation rates of a few percent per freefall time, as observed in most real clouds. Simulation movies available: \url{http://www.mso.anu.edu.au/~chfeder/pubs/ineff_sf/ineff_sf.html}.}
\label{fig:f15}
\end{figure*}

Figure~\ref{fig:f15} shows four simulations by that quantify the effects of gravity alone (top-left panel), added turbulence (top-right panel), added magnetic fields (bottom-left panel), and added jet/outflow feedback (bottom-right panel). With gravity alone, the star formation rate proceeds close to the maximum freefall rate. Turbulence, magnetic fields, and finally feedback, reduce the star formation rate by a factor of 2--3 in each step, leading to near-realistic (observed) typical star formation rates of a few percent per freefall time. Typical observed values are around 1--2\% per freefall time \citep{KrumholzTan2007,Federrath2013sflaw,OnusKrumholzFederrath2018}. The best values achieved in the most complex simulation is a star formation rate per freefall time of $\sim4\%$, which still overestimates the star formation rate by a factor of $\sim2$, most likely because radiation feedback was not included in these simulations, which will also have a profound impact on the initial mass function of stars \citep{OffnerEtAl2009,Bate2012,KrumholzKleinMcKee2012,FederrathKrumholzHopkins2017,CunninghamEtAl2018,GuszejnovEtAl2018}.

\section{The role of turbulence}

\begin{figure*}
\centerline{\includegraphics[width=1.0\linewidth]{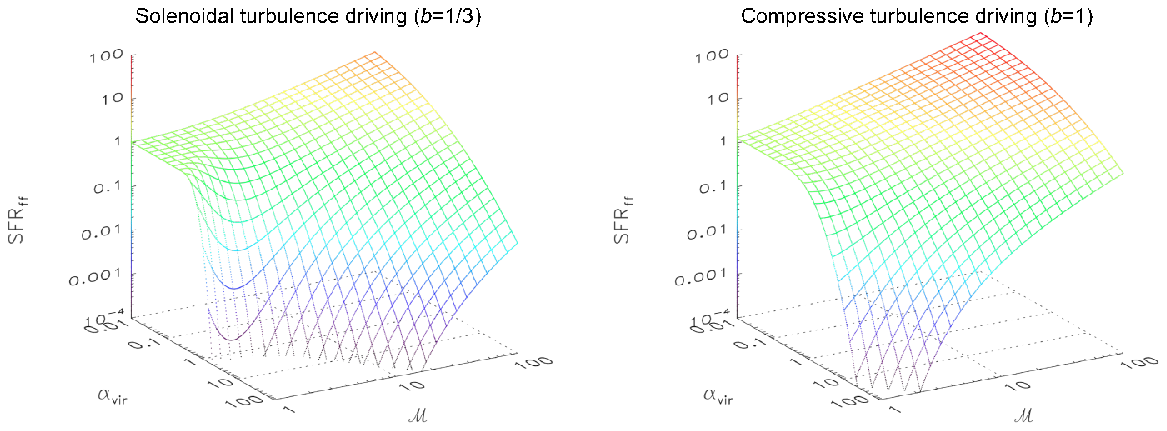}}
\caption{Theoretical predictions of the star formation rate per freefall time as a function of virial parameter $\alpha_\mathrm{vir}$ and turbulence Mach number $\mathcal{M}$, for solenoidal driving of the turbulence ($b=1/3$; left-hand panel) and compressive driving ($b=1$, right-hand panel). The difference in star formation rate is about an order of magnitude in the typical parameter range of observed clouds ($\alpha_\mathrm{vir}\sim1$, $\mathcal{M}\sim10$) between the two driving modes of turbulence, emphasizing the importance of the turbulence driving mechanism in controlling the formation of stars. Figure adopted from \citet{FederrathKlessen2012}}
\label{fig:fk12solcomp}
\end{figure*}

Recent theoretical advances allow us to predict the star formation rate based upon four fundamental physical parameters of a molecular cloud \citep{KrumholzMcKee2005,PadoanNordlund2011,HennebelleChabrier2011,FederrathKlessen2012},

\vspace{0.1cm}
\begin{enumerate}[labelwidth=0.6cm,labelindent=10pt,leftmargin=0.7cm,label=\bfseries \arabic*.,align=left]
\setlength{\itemsep}{5pt}

\item the virial parameter $\alpha_\mathrm{vir}=2E_\mathrm{turb}/E_\mathrm{grav}$ (ratio of turbulent kinetic to gravitational energy),

\item the sonic Mach number $\mathcal{M}=\sigma_v/c_\mathrm{s}$ (ratio of turbulent velocity dispersion to sound speed),

\item the turbulence driving parameter $b$, i.e., whether the turbulence is driven solenoidally ($b=1/3$) or compressively ($b=1$) \citep{FederrathKlessenSchmidt2008,FederrathDuvalKlessenSchmidtMacLow2010}, and

\item the plasma $\beta = P_\mathrm{th}/P_\mathrm{mag}$ (ratio of thermal to magnetic pressure).

\end{enumerate}
\vspace{0.1cm}

These theories rest on the statistics of supersonic magnetized turbulence, its density probability distribution function and power spectrum \citep{Federrath2013}. Figure~\ref{fig:fk12solcomp} shows the theoretical predictions of the star formation rate per freefall time as a function of $\alpha_\mathrm{vir}$ and $\mathcal{M}$, for solenoidal driving (left-hand panel) and compressive driving (right-hand panel). These can be understood as follows: 1) Large $\alpha_\mathrm{vir}>1$ correspond to unbound clouds, such that the star formation rate decreases very quickly with increasing $\alpha_\mathrm{vir}$. Typical clouds have $\alpha_\mathrm{vir}\sim1$ \citep{KauffmannPillaiGoldsmith2013,HernandezTan2015}. 2) Increasing $\mathcal{M}$ produces stronger shocks, and hence a higher dense-gas fraction \citep{KonstandinEtAl2012apj,KainulainenFederrathHenning2013,KainulainenFederrathHenning2014}. 3) More compressive driving of the turbulence also leads to enhanced dense-gas fractions compared to solenoidal driving \citep{FederrathKlessenSchmidt2008,FederrathDuvalKlessenSchmidtMacLow2010}. The latter effect can be very strong, leading to star formation rates that differ by more than a factor of 10 between solenoidal and compressive driving, as confirmed in simulations \citep{FederrathKlessen2012}.

\begin{figure*}
\centerline{\includegraphics[width=0.75\linewidth]{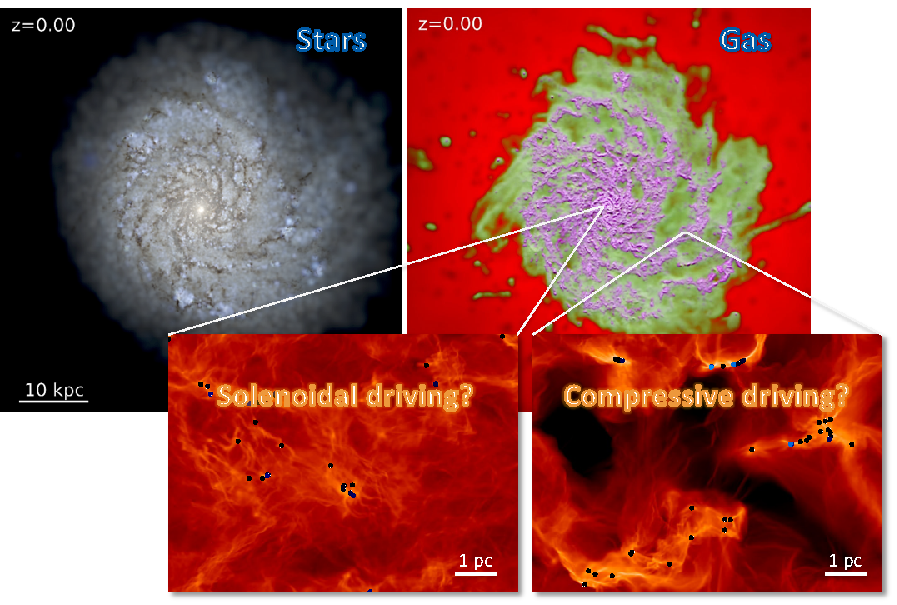}}
\caption{Top panels: stars and gas in a Milky Way type galaxy formed in the FIRE simulations \citep{HopkinsEtAl2014}. Bottom panels: two high-resolution idealized molecular cloud simulations focussing on small scales \citep{FederrathKlessen2012,FederrathKlessen2013}. Currently the galaxy-scale (top) and cloud-scale (bottom) simulations are completely decoupled from one another. By combining them we can determine the dominant turbulence driving mechanisms in different environments. For example, the central molecular regions of galaxies may be subject to strong shearing motions and hence may be dominated by solenoidal driving of turbulence, while spiral-arm clouds may form via cloud-cloud collisions and spiral-arm compression, which are in the category of compressive drivers.}
\label{fig:fire}
\end{figure*}

Given the strong effect of turbulence driving on star formation, a logical next step is to find out which driving mode (solenoidal versus compressive) dominates in different regions of the Galaxy. Figure~\ref{fig:fire} shows a schematic to illustrate the potentially important effect of environment. For example, spiral-arm compression and cloud-cloud collisions in the galactic disk may be considered compressive drivers, while shear in the Central Molecular Zone is a solenoidal driver \citep{FederrathEtAl2016,FederrathEtAl2017iaus}. Determining the dominant turbulent driving mode in different clouds and galactic environments is the combined effort of observers and theorists \citep{BruntFederrath2014,JinEtAl2017,KoertgenFederrathBanerjee2017,OrkiszEtAl2017}.

\section{The role of magnetic fields and feedback}

\begin{figure*}
\centerline{\includegraphics[width=1.0\linewidth]{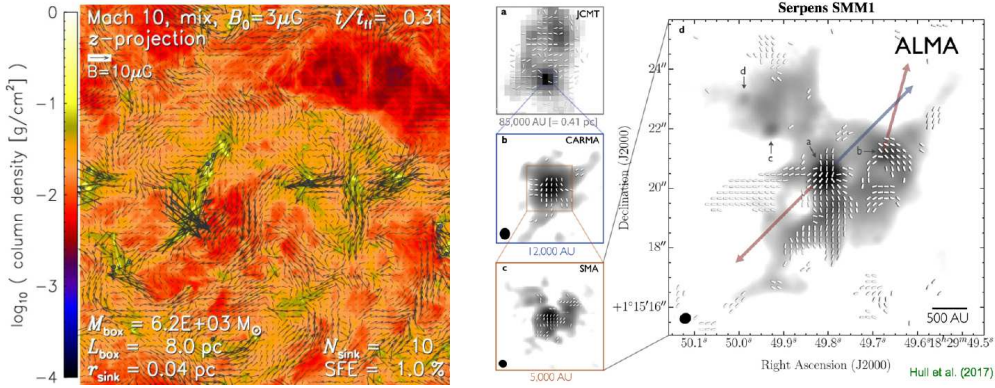}}
\caption{Left: Numerical simulation of a magnetized cloud with magnetic field vectors superimposed on the column density. Right: Zoom-in on a dense core in Serpens, observed with JCMT, CARMA, SMA, and ALMA, probing the magnetic field structure via dust polarization. The cloud breaks up into several individual over-densities (labelled a--d) and jets/outflow may be launched from of these regions (indicated by the arrows). The magnetic field structure in both simulations and observations is highly complex, with a combination of ordered and turbulent field configurations. Images adopted from \citet{FederrathKlessen2012} and \citet{HullEtAl2017ALMA}.}
\label{fig:fk12b}
\end{figure*}

The simulations and theoretical models discussed in the previous section have shown that the magnetic field can reduce the star formation rate by a factor of 2--3 compared to purely hydrodynamic cases. Here we want to focus on the effects of the magnetic field configuration. Recent observations of dust polarization with ALMA reveal highly complex magnetic field structures including some ordered, chaotic (turbulent), and rotational components \citep{HullEtAl2017ALMA,LeeChinFeiEtAl2018,ZhangYichenEtAl2018,CoxEtAl2018}. Many of these cases differ substantially from the classical hour-glass morphology. The most likely reason for this is the turbulence in the parental molecular cloud, creating flows that twist, tangle and compress the magnetic field lines into equally complex structures as the velocity and density fields of these clouds and cores \citep{Federrath2016jpp}. Figure~\ref{fig:fk12b} shows an example of the complexity of the magnetic field in both simulations and observations.

\begin{figure*}
\centerline{\includegraphics[width=1.0\linewidth]{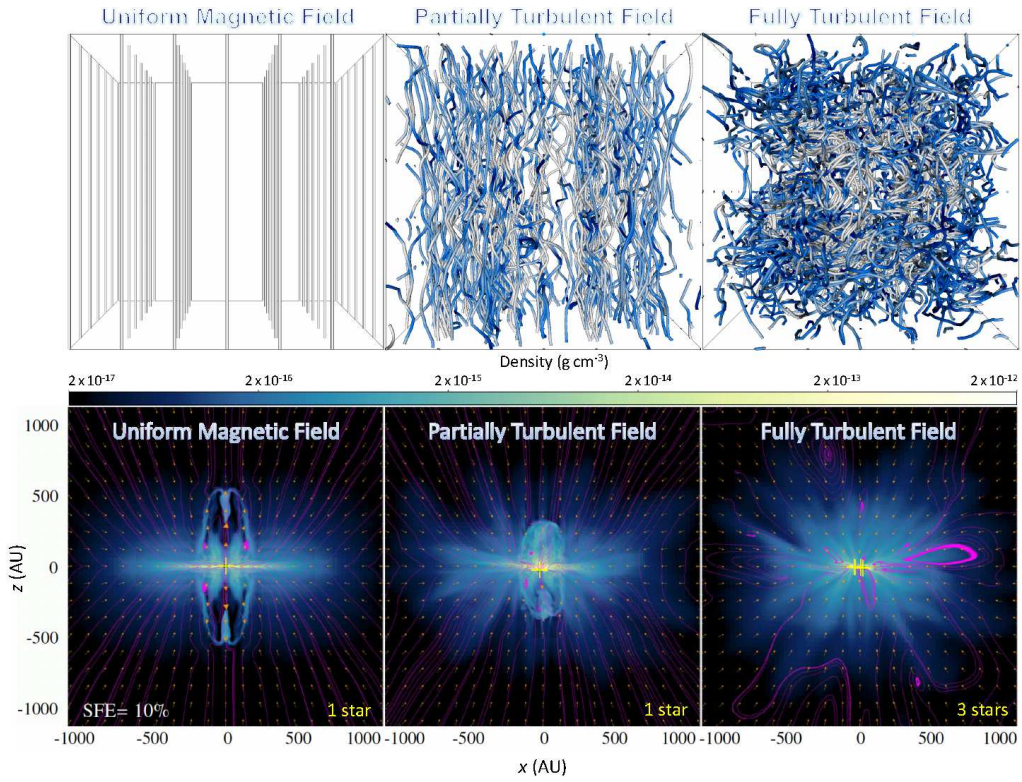}}
\caption{Three simulations of solar-mass star formation. The simulations are identical, except for the magnetic field structure at the start of the simulations. The initial magnetic field configuration is shown in the top panels; from left to right: 1) uniform, 2) partially turbulent, 3) fully turbulent. The bottom panels show the respective accretion disks that form in these three different configurations, when the star(s) has/have accreted $0.1$ solar masses. The star(s) (shown as crosses) formed in the disks, seen edge-on in the images. The magnetic field configuration has a strong impact on the outflow/jet strengths (with the uniform and partially turbulent runs launching outflows, while no jets are produced in the fully turbulent case) and on the fragmentation of the core/disk (with the fully turbulent field leading to the formation of 3 stars as opposed to only 1 star each, in the uniform and partially turbulent simulations). Images adopted from \citet{GerrardFederrathKuruwita2019}.}
\label{fig:gfk19}
\end{figure*}

The magnetic field configuration may have profound consequences for star formation, the structure of the disks, the launching of jets and outflows, and ultimately for the formation of planets. Figure~\ref{fig:gfk19} shows recent simulation results that aim to isolate the effect of the magnetic field structure on the formation of solar-mass stars. Three identical simulations are compared, only differing in the initial field configuration. The panels from left to right show simulations with initial magnetic field in a uniform (ordered) configuration (aligned with the rotation axis of the core), a partially turbulent case (in which the ordered and turbulent components have the same strength), and a fully turbulent field. The total strength of the initial magnetic field was kept constant at $100\,\mu\mathrm{Gauss}$ in all three cases, as were all the other initial parameters \citep{GerrardFederrathKuruwita2019}. The bottom panels of Figure~\ref{fig:gfk19} show the gas density when the protostar(s) have reached $0.1$ solar masses. Most importantly, we see that the uniform and partially turbulent cases launch jets and outflows along the rotation axis of their disks, while the fully turbulent magnetic field is not capable of launching jets at all. The turbulent field also leads to the formation of 3 stars compared to only a single star in both uniform and partially turbulent field configurations. In summary, the magnetic field structure strongly affects the disk evolution, outflow efficiency and fragmentation, and thus, the mass distribution of stars.

\section{Summary and conclusions}

The main conclusions are

\vspace{0.1cm}
\begin{itemize}[itemindent=0.2cm]
\setlength{\itemsep}{5pt}

\item \quad Star formation is inefficient. The relevant physical processes opposing fast gravitational collapse and making star formation slow and inefficient are: 1) turbulence, 2) magnetic fields, and 3) feedback. Each of these processes reduces the star formation rate by factors of 2--3. In their combination, turbulence, magnetic fields and feedback can yield realistic, observed, low star formation rates per freefall time, of only a few percent.

\item \quad Turbulence can reduce the star formation rate by more than a factor of 10 when driven by a solenoidal driver (such as shear), compared to a compressive driver (such as supernova explosions or galactic spiral-arm compression). Solenoidal driving may dominate in the Central Molecular Zone, while compressive drivers may be more relevant in spiral-arm clouds. Determining the mixture of turbulent modes is critical to understand and predict star formation.

\item \quad ALMA observations can now probe the magnetic fields in dense protostellar cores down to scales of a few astronomical units. They reveal highly complex field geometries, containing both ordered and turbulent magnetic field components. Simulations with ordered, partially turbulent and completely turbulent fields show that the outflows and jets launched and the stars formed depend significantly on the different field configurations in these dense cores. More observational constraints and simulations that take these complex magnetic field structures into account will be needed to unravel the initial mass function of stars and the origin of binaries and planets.

\end{itemize}

\section*{Acknowledgments}

C.~F.~acknowledges funding by the Australian Research Council (Discovery Projects DP170100603 and Future Fellowship FT180100495), and the Australia-Germany Joint Research Cooperation Scheme (UA-DAAD). We further acknowledge high-performance computing resources provided by the Leibniz Rechenzentrum and the Gauss Centre for Supercomputing (grants~pr32lo, pr48pi and GCS Large-scale project~10391), the Partnership for Advanced Computing in Europe (PRACE grant pr89mu), the Australian National Computational Infrastructure (grant~ek9), and the Pawsey Supercomputing Centre with funding from the Australian Government and the Government of Western Australia, in the framework of the National Computational Merit Allocation Scheme and the ANU Allocation Scheme.
The simulation software FLASH was in part developed by the DOE-supported Flash Center for Computational Science at the University of Chicago.

\begin{discussion}

\discuss{Cuntz}{Please comment on the effect(s) of metallicity, as the latter is expected to impact the cooling processes.}
\discuss{Federrath}{Indeed, the composition (metallicity) determines the thermodynamic response of the gas. This is particularly important when comparing present-day star formation with star formation in the early Universe, i.e., for the formation of the First Stars, where cooling was not as efficient as in solar-metallicity gas.}

\discuss{Khaibrakhmanov}{What is the physical mechanism that determines filament formation at the sonic scale?}
\discuss{Federrath}{Filaments and cores may form at the sonic scale, because it is the scale, where shocks transition from supersonic to subsonic speeds. Thus, filaments may be the stagnation points of compressive turbulent flows, i.e., filaments forming at the intersection of planar shock waves.}

\discuss{Linsky}{Can you explain the initial mass function of stars? Was Salpeter correct in his prediction of the IMF?}
\discuss{Federrath}{The IMF is one of the biggest challenges to explain and understand. It is likely determined by a combination of physical processes, in particular gravity, turbulence, magnetic fields, and feedback (by jets/outflows and radiation). The IMF will be the focus of advanced simulations in the next few years.}

\end{discussion}

\end{document}